\documentclass[useAMS,usenatbib,usegraphicx,onecolumn]{mn2e} 

\title[NIRS0S review]{Overview of the Near-IR S0 galaxy Survey (NIRS0S)}

\author[E-mail:eija.laurikainen@oulu.fi]{E. Laurikainen$^{1,2}$\thanks{E-mail:eija.laurikainen@oulu.fi}, H. Salo$^{1}$, R. Buta$^{3}$, J. H. Knapen$^{4,5}$\\
$^{1}$Dept. of Physics/Astronomy Division, University of Oulu, FI-90014 Finland\\
$^{2}$Finnish Centre of Astronomy with ESO (FINCA), University of Turku, V\"ais\"al\"antie 20, FI-21500, Piikki\"o, Finland\\ 
$^{3}$Department of Physics and Astronomy, University of Alabama, Box 870324, Tuscaloosa, AL 35487\\
$^{4}$Instituto de Astrof\'isica de Canarias, E-38200 La Laguna, Tenerife, Spain\\
$^{5}$Departamento de Astrof\'isica, Universidad de La Laguna, E-38205 La Laguna, Tenerife, Spain}
\begin{document}

\date{Accepted to Advances in Astronomy, Nov 28, 2011}


\maketitle

\label{firstpage}
 
\begin{abstract}
An overview of the results of the Near-IR S0 galaxy Survey (NIRS0S) is
presented. NIRS0S is a magnitude (m$_B$$\leq$12.5 mag) and inclination
($<$ 65$^o$) limited sample of $\sim$ 200 nearby galaxies, mainly S0s,
but include also Sa and E galaxies. It uses deep $K_s$-band images,
typically reaching a surface brightness of 23.5 mag
arcsec$^{-2}$. Detailed visual and photometric classifications were
made, for the first time coding also the lenses in a systematic
manner. The main analysis methods include 2D multi-component
decomposition approach, and Fourier analysis of the non-axisymmetric
structures.  As a comparison sample, a similar sized spiral galaxy
sample with similar image quality was used.  The main emphasis were to
study whether the S0s are former spirals in which star formation has
been ceased, and also, how robust are bars in galaxies. Based on our
analysis the Hubble sequence was revisited: following the early idea
by van den Bergh we suggested that the S0s are spread throughout the
Hubble sequence in parallel tuning forks as spirals (S0a, S0b, S0c
etc.). This is evidenced by our improved bulge-to-total ($B/T$) flux
ratios in the S0s, covering the full $B/T$ range, reaching small
values typical to late-type spirals. The properties of bulges and
disks in S0s were found to be similar to those in spirals and also, the
masses and scale parameters of the bulges and disks to be
coupled. It was estimated that the spiral bulges brighter than -20 mag
in $K$-band are massive enough to be converted into the bulges of S0s
merely by star formation. Bars were found to be fairly robust both in
S0s and spirals, but inspite of that bars might evolve significantly
within the Hubble sequence.

\end{abstract}


\section{Introduction}

We review the main results of the Near-IR S0 galaxy Survey (NIRS0S)
obtained so far \footnote{NIRS0S was performed in collaboration with
  the Universities of Oulu (Finland; Laurikainen $\&$ Salo),
  Alabama (USA; Buta), and Instituto de Astrof\'isica de Canarias
  (Spain; Knapen). The project was allocated a considerable amount of
  observing time at ESO (NTT, Chile), WHT, NOT, TNG (La Palma), CTIO
  (La Serena) and FLMN (Arizona) during 2004-2009.}. NIRS0S is a
magnitude (m$_B$$\leq$12.5 mag) and inclination (less than 65$^o$) limited
sample of $\sim$ 200 nearby
galaxies, mainly S0s, but include also Sa spirals and 25 late-type
ellipticals. Late-type ellipticals were included for not to miss any
potentially misclassified S0s. The observations were done in
the $K_s$-band, carried out using 3-4 meter sized ground-based telescopes with
sub-arcsecond pixel resolution. The images are deep, typically
reaching a surface brightnesses of 23.5 mag arcsec$^{-2}$ in azimuthally
averaged profiles ($\sim$ 2
mag deeper than the 2MASS images), thus allowing the detection of the
faint outer disks in S0s. Our main emphasis was to address possible
secular evolutionary processes in galaxies by comparing the photometric
properties of S0s and spirals, based on similarly selected
samples, with similar image quality.  

In the early classification by Hubble [37], the S0s were an enigmatic
group of galaxies between the ellipticals and early-type spirals, and
since then they have appeared to be important in any galaxy
evolutionary model. S0s are suggested to evolve from spirals, either
by internal secular processes in isolation, or by gas stripping
mechanisms, followed by quiescent star formation ([46]): in cluster
environment by ram pressure stripping ([36]) or by galaxy harassment
([57]), and in galaxy groups by tidal encounters (e.g. [38]).
Alternatively, S0s have been suggested to form by galaxy mergers in a
similar manner as elliptical galaxies ([68],[6]), or have accreted
most of their mass in minor mergers ([3], [41]).

Originally, the stripping scenario was adopted because S0s were found
to appear mainly in galaxy clusters ([26]), and also because they were
kinematically more related to spirals than to ellipticals
([27]). However, the large bulge fractions found by Simien $\&$ de
Vaucouleurs ([67]) in S0s has worked as a counterargument to this
scenario, because spiral galaxies do not have sufficient amount of
interstellar gas to build such large bulges ([45]). Some S0s are also
found to have chemically and kinematically decoupled cores ([1]),
which are generally interpreted in terms of sinking of small gas poor
satellites into the main galaxies.  Likewise, the merger approach has
problems, at least if interpreted as the dominant mechanism for
producing the S0s.  For example, modern semi-analytical merger models
(e.g. [41]) successfully produce the large bulge masses by Simien $\&$
de Vaucouleurs [67], but fail to recover the observed trend in
$\alpha$/Fe elements with stellar velocity dispersion ([69],
[40]). They have also difficulties to explain the observed tight
luminosity-size relation for the S0s ([60]).

The morphological division between E and S0 galaxies was obscured by
the detection of boxy and disky ellipticals by Bender ([12]; see also
[44]).  It was also found that both the galaxy luminosity and the
degree of rotational support ($\sigma/V_{max}$) correlate with the
isophotal shape of the elliptical galaxies ([13]).  Among other
things, these findings led King ([42]) and Djorgovski ([25]) to
announce that the Hubble sequence is breaking down and should be
replaced by a more physical approach, based on measured physical
parameters, related to galaxy kinematics. Recently it was suggested by
Cappellari et al. [22] that such a parameter could be $\lambda_{Re}$,
a proxy for the specific angular momentum in galaxies. Based on
$\lambda_{Re}$ they showed that most S0s, and also even 2/3 of the
elliptical galaxies in the nearby Universe are rotationally
supported. It was also shown by Emsellem et al. [29] that this
kinematic parameter has no clear correlation with the boxy/disky
isophotal shapes.  These findings are suggested to put into a new
light many of the previous results, including the morphological
classification of the early-type galaxies ([22]).  NIRS0S has a large
overlap with the sample by Cappellari et al. [22] (used in [29]), thus
allowing to compare the kinematic classification with the new
morphological classification of S0s made by Laurikainen et al. ([54]).

In the following an overview of the main results of NIRS0S is given,
discussed in conjunction of some recent studies of S0s (more complete
references can be found in the original papers).  As a comparison
sample, the Ohio State University Bright Spiral Galaxy Survey
(OSUBSGS; [31]) has been used, for which galaxies we have carried out
similar analysis as for NIRS0S. Our main analysis methods include
detailed visual and photometric classification of galaxies, 2D
multi-component structural decompositions, Fourier analysis for
calculating the properties of bars, and isophotal analysis.  In this
paper we also analyze a sub-sample of NIRS0S, which have kinematic
information given by Emsellem et al. ([29]). A picture is outlined
supporting the view that S0s in general are evolved from spiral
progenitors. They are suggested to be spread throughout the Hubble
sequence forming S0a, S0b, S0c types, in a similar manner as spirals.
Our scenario, originally outlined in Laurikainen et al. [53], is
  similar to that suggested recently by Kormendy $\&$ Bender [88].

\section{NIRS0S atlas}

NIRS0S Atlas ([54]) presents the flux calibrated
images, shown in a logarithmic scale.
Also shown are the dimensions of the identified structures, as well as
the radial profiles of the isophotal analysis results (ellipticity,
position angle, and the parameter b4, which measures the isophote's
deviations from perfect ellipticity).  An example of the layout of the
atlas images is shown in Figure 1.

\subsection{Morphological classification and how it compares with kinematic classification}
 
The classification is based on de Vaucouleurs' revised Hubble-Sandage
system ([73], see also [19], [21]), which includes the stage (S0$^-$,
S0$^o$, S0$^+$, Sa), family (SA, SAB, SB), variety (r, rs, s),
outer-ring or pseudo-ring (R, R'), possible spindle shape, and the
presence of peculiarity. The ring classification follows Buta $\&$
Crocker ([15]) and Buta ([16]). What is new in our classification is
that lenses (nuclear, inner, outer) are, for the first time,
systematically coded for a significant sample of S0s.  A new lens
type, 'barlens', is also introduced: it is a lens-like structure
embedded in a bar, with a fairly sharply declining outer surface
brightness profile - this component has been often erroneously mixed
with the bulge. Barlenses are shorter than the main bars, and are
typically identified inside strong bars. Due to the sub-arcsecond
pixel resolution it was also possible to classify the central
structures like nuclear bars, rings and lenses in a systematic manner.

Besides the above discussed visual classification, we also made a 
photometric classification where faint structures are
identified even if they were not directly visible in the images. Faint
structures, overshadowed by bulges, were detected after subtracting
the bulge model in the decomposition, and/or by making unsharp
masking. Lenses were detected by inspecting the surface brightness
profiles, in which they generally appear as nearly exponential
subsections, however with more shallow profiles than the main
disk component. Photometric classification turned out to be critical
for separating S0s from the elliptical galaxies at the E/S0 interface.
 
NIRS0S has 66 galaxies in common with the 3DAtlas, which is a
volume-limited sample of 260 early-type galaxies, with the kinematic
$\lambda_{Re}$ parameter available ([29]).  $\lambda_{Re}$ is a proxy
for the specific angular momentum in galaxies, and can be used as a
measure for the rate of the kinematic support, thus dividing the
galaxies to fast and slow rotators. It is measured within one
effective radius of galaxy brightness thus covering well the bulge
region. Of the 66 galaxies in common, 9 were ellipticals, and 56 were
S0-S0/a galaxies in the Third Reference Catalog of Bright Galaxies
([73], RC3). However, in our classification only 2 out of the 9
galaxies were truly ellipticals. Kinematically, one of the two
ellipticals is a slow rotator, whereas the other is a fast rotator
showing an inner disk. The complete NIRS0S has 26 galaxies classified
as ellipticals in RC3, of which only 3 remained ellipticals in our
photometric classification. This comparison shows that generally the
Hubble stage, based on the morphological and kinematic classification
is consistent with each other (notice that NIRS0S does not include
highly inclined galaxies).  However, based on $\lambda_{Re}$ not all
galaxies with prominent outer disks are fast rotators.  Slow rotators
include NGC 4552, NGC 5846, NGC 5631 and NGC 6703, in which the S0
nature is confirmed by the detection of a lens (NGC 5846 even has
multiple lenses). Slow rotators comprise also NGC 4406 and NGC 4472,
which have exponential outer profiles, used as evidence of a disk in
our photometric classification. In conclusion, this means that
  $\lambda_{Re}$ alone cannot be used to distinguish the Hubble stage
  among the S0s and the elliptical galaxies.

\section{Properties of bars}

\subsection{Morphology of bars in S0s}

Bars are the main characteristic common in S0s and spirals.  They can
be regular elongated structures (NGC 4608), have condensations at the
two ends of the bar (ansae; e.g. NGC 2983), or show X-shaped structure
in the inner part of the bar (e.g. IC 5240) (Fig. 2).  Besides ansae,
NGC 2983 and NGC 4608 have also barlenses. Both structures are
illustrated in Figure 2 by subtracting the underlying bar+bulge model
from the image of NGC 2983: the residual image shows a nearly
spherical barlens, and two blops (ansae) that form part of the bar.

It was shown by Laurikainen et al. ([51]) that ansae in bars are
characteristic to S0s, appearing less frequently in spiral galaxies
(40$\%$ and 12$\%$, respectively). This was shown simultaneously also
by Martinez-Valpuesta, Knapen $\&$ Buta ([55]), who additionally suggested that
strong bars (B) have ansae more frequently than weak bars (AB). Both
studies used a sub-sample of NIRS0S, and in [55]
also the de Vaucouleurs' atlas by Buta et al. ([19]).  By
repeating the statistics for the complete NIRS0S sample in this study,
using the classifications of the NIRS0S Atlas, we find ansae in 30$\%$
of the S0s, which is still clearly higher than in spirals. However,
now the fraction of ansae is the same for strong (B) and weak (AB)
bars. In Laurikainen et al. [52] the large fraction of ansae bars
among the S0s was discussed as evidence of bar evolution in the Hubble
sequence.

In NIRS0S Atlas X-shaped bars appear in 9 galaxies, which is 8$\%$ of
all barred galaxies ([54]).  This is important, because all the
galaxies in NIRS0S have inclinations less than 65$^0$, although the
X-shaped bars are generally discussed in the context of nearly edge-on
galaxies, suggested to be formed by vertical thickening due to
buckling effects ([23], [79]). It was shown by Athanassoula $\&$
Beaton [7] that the vertical thickening might still be recognized at
inclinations of 77.5$^0$, whereas our study shows that the X-shaped
bar morphology can be visible even in almost face-on galaxies. A nice
example of such a bar appears in IC 5240 (Fig. 2), which has an
inclination of 49$^0$ (taken from [54]).  Evidently, bar buckling
alone, leading to a vertical thickening of a bar, is not sufficient to
explain such bar morphologies.

Multiple bars appear in $\sim$ 20$\%$ of the S0s [52], which is a
similar number as found previously also for spirals [75, 76].  It is
worth noticing that S0/a galaxies have a larger number of multiple
bars [52]. A prototypical case where two bars appear nearly
perpendicular to each other is shown in Figure. 3, in which both the 
nuclear and the main bar are surrounded by lenses.  Some galaxies,
like NGC 2681, have even three bars.

 \subsection{Fourier properties of bars: S0s vs. spirals}

The properties of bars in S0s (using a sub-sample of NIRS0S) and
spirals (using OSUBSGS) has been studied by Laurikainen et al. [51]
using a Fourier method.  For the complete NIRS0S they will be
presented in a forthcoming paper.  Tangential forces induced by the
non-axisymmetric structures, mainly bars, were measured using a bar
torque approach ([24]), applying the polar method by Salo et
al. ([62], [64]; see also [47]). In this method the surface brightness
at each radius is decomposed into azimuthal Fourier components. Using
the even Fourier components (typically up to $m$=10), the
gravitational potential in the equatorial plane of the galaxy is
obtained by FFT over azimuth, combined with direct summation over
radial and vertical coordinates.  An important advantage of the polar
method is the suppression of spurious force maxima which may arise in
the direct 3D Cartesian FFT integration in the noisy outer disks. The
calculated Fourier amplitude profiles also provide another useful tool
for characterizing bars (see below).
 
As a measure of bar strength, we use $Q_g = \max({F_T/<F_R>})$, the
maximum of tangential force amplitude relative to the mean
axisymmetric radial force, evaluated at the region of the bar.  The
mass-to-light ratio ($M/L$) is assumed to be constant, and the
vertical scale height of the disk (and bar) is estimated from the
exponential scale length, using a Hubble type dependent mean ratio.
Also, the different 3D density distribution of the bulge is corrected,
based on bulge models obtained from decompositions.  The effect of
including dark halo force field was also investigated, but its
influence on $Q_g$ appeared to be insignificant at the bar region
([63], [17]). These calculations also give a proxy for the bar length,
$rQ_g$, which is the radius where the maximum tangential force $Q_g$
occurs. Bar lengths were estimated also from the phases of the
  $A_2$ Fourier amplitude by assuming that it is maintained nearly
  constant in the bar region (a correlation between this bar length
  and $rQ_g$ was shown by [48]). The maximum of $m$=2 Fourier
amplitude, $A_2$, was used as an estimate of the relative brightness
of the bar. These properties are shown in Figure 5 in Laurikainen et
al. [51] as a function of Hubble: bars grow in length and in relative
brightness ($A_2$) towards the early-type galaxies, but for $Q_g$ the
trend is opposite. Notice that although the bar ellipticity
  (shown in the same figure) correlates with $Q_g$ ([91]), it has no
  systematic correlation with the Hubble type. In Laurikainen et
el. [48 ] the tendency of weakening bar strengths ($Q_g$) towards the
early-type galaxies was explained by a dilution effect due to the
more massive bulges in the early-type galaxies (e.g. the average $Q_g$
parameter may decrease even if the average $A_2$ amplitude increases,
since the bulge contribution to radial force becomes more important
toward earlier types). There exist a correlation also between
  $Q_g$ and $A_2$, but for the above reason the correlations are
  different for the early and late-type galaxies (see Fig. 8 in
  [48]).  The obtained tendency for bar lengths was originally shown
by Elmegreen $\&$ Elmegreen [28].

For a sub-sample of 26 barred galaxies in NIRS0S, the radial $A_2$
profiles were fitted by single (SG) and double (DG) Gaussian functions
by Buta et al. [18]. It appeared that 65$\%$ of the bars in S0-S0/a
galaxies have single Gaussian profiles, whereas 35$\%$ are best fitted
by two Gaussian functions. Typical examples of such profiles are shown
in Figure 4. The galaxies with DG bars typically have also significant
higher Fourier modes in the bar region ($m$=4, 6, 8), in addition to
$m$=2 ([51]).  It was discussed by Buta et al. that the DG-profiles
are similar to those predicted by the simulation models (e.g., [5]) in
which the bar transfers a large amount of angular momentum to the
halo.  An attempt to associate the DG-profiles to specific
morphological structures was made by Laurikainen et al. [51] who
suggested that the fat or double-peaked Gaussian amplitude profiles
are due to two bar components, a long and narrow bar, and a shorter
component in the inner parts of the bar (or an inner oval). DG
  bars were found to be more prominent, not only in terms of $Q_g$,
  but also in $A_2$ and bar length [51]. In Laurikainen et al. [54]
these inner bar components were associated mainly with barlenses
(though some of them can be ovals), which are found to appear in
30$\%$ of barred S0-S0/a galaxies in NIRS0S. A good example is NGC
4314 (Fig. 1), in which the barlens is the fat elongated structure
inside the bar.  Erwin et al. [30] have discussed two S0s, in which a
superposition of a classical and a pseudo-bulge was suggested. 
  These galaxies are NGC 2787 and NGC 3945, which form part of
  NIRS0S. In both galaxies the component interpreted as a pseudo-bulge
  by Erwin et al., is called as a fat inner bar component in [51], and
  more recently defined as a barlens by us [54].

 \subsection{Do S0s have the bar strengths expected for systems not accreting any gas?}

The above question was recently made by Buta et al. [86] with the main
emphasis to test the hypothesis by Bournaud $\&$ Combes [14], in which
multiple bar episodes are expected in the Hubble time. In this
scenario bars form and evolve in galaxies when they have gas, and the
evolution stops when the gas in used in star formation. These stars
are then transferred into the bulge, for example by bars or spiral
arms in the central regions of the galaxies.  When the central mass
concentration formed by star formation becomes very high, the bar will
be destroyed.  Therefore, if bar strength varies over time the
relative frequency of galaxies in each $Q_g$ bin tells us the relative
amount of time a galaxy spends in a certain bar state (strong, weak,
non-barred). For spirals this was first tested by Block et
al. [80]. They suggested that galaxies might have doubled their mass in
10$^{10}$ years (see Fig. 5, left panel), evidenced by the extended tail
towards strong bars, and the lack of weak bars, which features are
predicted in the strong gas accretion models by Bournaud $\&$ Combes
[14].

This test was later repeated by Buta, Laurikainen and Salo [87] for the same
galaxy sample, but using the refined bar torque method described in
the previous section (we also discussed why the obtained $Q_g$
distribution was different from that by Block et al. [80]).  In
  Buta et al. [17] the bar and spiral fluxes were additionally
  separated from each other. Although the correction affected $Q_g$ in
  a few individual cases having very strong spiral arms, it barely
  affected the $Q_g$ distribution. The refined $Q_g$ distribution
  ([87], [17]) has a large number of weak bars lacking from that
  obtained by Block et al.  Likewise, it has a slightly smaller number
  of very strong bars. In fact, the obtained $Q_g$ distribution (see
  Fig. 8a in [17]) largely resembles the non-accretion model by
  Bournaud $\&$ Combes shown in Figure 5 (left panel), thus supporting
  the view that bars in spirals are fairly robust. In Buta et al. [86]
  the $Q_g$ distribution for NIRS0S was calculated. Most importantly,
  a clear difference was found between S0s and early-type spirals (see
  Fig. 5, right panel).  This was suggested to support the view
  according to which S0s have not accreted gas for a long time,
  evidenced by the lack of the extended tail, and the existence of a
  large number of weak bars. As discussed above (see Section 3.2) the
  smaller number of strong bars among the S0s can be due to a dilution
  effect caused by the more massive bulges and thicker disks in
  S0s. However, it was also discussed by Buta et al. [86] that this
  cannot produce all of the difference in $Q_g$ between the S0s and
  early-type spirals: although spirals have a larger number of strong
  bars than S0s, still there are no galaxies having $Q_g$ $>$
  0.5. The conclusion in this study was that if S0s are
  stripped spirals, the weaker bars in S0s could indicate that
  bar evolution continues to proceed even after gas depletion in
  galaxies.

 \section{Lenses}       

Lenses appear as flat disk components with rather sharp outer edges
([43]).  However, not all lenses are directly visible in the
images. In the NIRS0S Atlas ([54]) lenses were generally detected as
exponential subsections in the surface brightness profiles.  NGC 524
(Fig. 6) shows all the main lens types, nuclear (nl), inner (l), and
outer lens (L). When the outer lens is very prominent compared to the
underlying disk, as in NGC 1533, it manifests as a broad bump in the
surface brightness profile, in this case having also some
characteristics of a ring (RL).

For clarity the examples shown of the different lens types are for
non-barred galaxies.  However, lenses appear both in barred (61$\%$)
and non-barred (38$\%$) S0s, based on the classification in NIRS0S
Atlas [54]. In Laurikainen et al. [52] even a larger fraction of
lenses was found, but it was based on a sub-sample of NIRS0S.  In
barred galaxies nuclear (nl) and inner (l) lenses typically end up to
the radius of the nuclear and the main bar, respectively, relating
them to resonances of the rotating bar (see NGC 1543 in Fig. 3).
However, not all lenses are related to resonances. For example, series
of lenses in some non-barred S0s appear, like in NGC 1411, which
cannot be immediately understood in the framework of the resonance
theory. Lenses can also be relics of significant star formation in the
spiral arms as suggested in the NIRS0S Atlas ([54]). Originally the
idea is from Bosma [8], who also confined the lens formation to the
epoch of galaxy formation. Lenses are also suggested to form by disk
instability, in a similar manner as bars [77].  It was further
suggested by Kormendy [43], that bars may gradually dissolve into
lenses. In fact there are many results in NIRS0S which are consistent
with this scenario: (a) lenses in barred S0s often end at the bar
radius, (b) S0s were found to have a smaller bar fraction, and a
larger fraction of lenses than spirals ([52]). Also, (c) dissolution
of bars would explain the large number of lenses in non-barred S0s in
a natural manner.  Using the ellipticity of a bar, a smaller bar
fraction in S0s, compared to that in spirals, was found also by
Aguerri, M\'endez-Abrey $\&$ Corsini [4].  Most probably lenses have
multiple origins, and in order to better understand their nature
detailed analysis of their dimensions and physical properties needs to
be performed.  A forthcoming NIRS0S paper will focus on that.

It is worth noticing that multiple lenses appear even in 25$\%$ of the
S0s in the NIRS0S Atlas, including barred and non-barred galaxies
[54], which needs to be understood in the formation and evolution of
galaxies. For example, if a large fraction of the mass in the S0s was
accreted by minor mergers, it needs to be understood how the
multiple lenses can survive through such processes.

\section{2D multi-component decompositions for NIRS0S}

\subsection{The multi-component approach}

A 2D multi-component code, BDBAR (written by Salo, and described in
[48], [49]), was used for decomposing the light
distributions of the $K_s$-band images into bulges, disks, bars, ovals
and lenses.  This multi-component approach turned out to be important,
not only for barred galaxies, but also for galaxies with prominent
lenses. Using artificial images this was tested by Laurikainen et
al. [50]. Figure 7a shows the surface brightness profile of a
synthetic image with a bulge and a disk, with random noise added, whereas 7b
and 7c show the same image after adding a small bar on top of
that. Making a simple bulge-disk decomposition for the barred
synthetic image (in the middle) overestimates $B/T$ (=0.36), due to
erroneous assignment of the bar flux to the bulge, whereas the
bulge-disk-bar decomposition (right) recovers the correct $B/T$ value
($B/T$=0.27). The residual image also shows a bar in the simple
bulge-disk model, but not in the bulge-disk-bar decomposition.

A test for the observed NIRS0S images was made by Laurikainen, Salo $\&$ Buta
[49], collected to Table 1, where 1D (using azimuthally averaged
profiles) and 2D decompositions are also compared.  It appears that
simple bulge-disk decompositions give a similar mean $B/T$-ratio,
independent of whether 1D or 2D fitting is used, whereas the
three-component approach gives significantly lower $B/T$. The value
$<B/T_K>$=0.55 in the bulge-disk decomposition is very similar
to $<B/T_B>$=0.57 as obtained by Simien $\&$ de Vaucouleurs [67]. We also
estimated that the different wavelengths used does not cause this
difference. Adding even more components (=''final'' model in the Table),
like nuclear bars, further lowers the $B/T$, but the change is not as
dramatic as between the 2 and 3 component models. The S\'ersic index
is also smaller in the bulge-disk-bar decompositions, but even in the
simplest models the mean value is not as large as 4, as often produced 
by merger simulations.

\begin{table}
\begin{center}
\caption{Mean $B/T$-ratio and S\'ersic $n$ parameter of the bulge as estimated using the 1D and 2D-decompositions,
applied for the $K_s$-band images. 
This table was originally published in Laurikainen et al. [49]. \label{tbl-1}}
\begin{tabular}{lcc}
                      &                 &                          \\
                      &                 &                 \\
method                &  $<B/T>$        & $<n_{bulge}>$  \\
                      &                 &                \\
\noalign{\smallskip}

2D (final)           & 0.25 $\pm$ 0.03  & 2.1 $\pm$ 0.1 \\ 
2D (bulge/disk/bar)  & 0.30 $\pm$ 0.03  & 2.1 $\pm$ 0.1  \\
2D (bulge/disk)      & 0.55 $\pm$ 0.06  & 2.6 $\pm$ 0.1  \\
1D                   & 0.48 $\pm$ 0.04  & 2.7 $\pm$ 0.4   \\
                     &                  &                 \\
\end{tabular}
\end{center}
\end{table}

\subsection{Low B/T values obtained both for S0s and spirals}

The 2D multi-component decompositions for the complete NIRS0S sample
are discussed by Laurikainen et al. [53], where they are also compared
with a similar sized sample of OSUBSGS spirals, using the same
decomposition approach (see [48], [51]). Internal dust correction was
applied to all galaxies, in a similar manner as in Graham $\&$ Worley
[35].  For spirals these results are in good agreement with the
bulge-disk decompositions made for non-barred galaxies ([78,84]), and
with other multi-component decompositions ([81],[82],[83]). It is
encouraging that Weinzirl et al. [81], applying bulge-disk-bar
decompositions, obtained very similar low $B/T$ ratios as obtained
previously by Laurikainen et al. ([48],[51]) for the same sample of
spirals. In both studies $M/L$ ratio was assumed to be radially
constant. Graham $\&$ Worley [35] also obtained fairly low
$B/T$ values, but their study uses heterogeneous data from the
literature, including simple and multi-component decompositions.
\footnote{The 2-component decompositions are thoroughly discussed in the
  recent review by Graham, to appear in the next edition of ``Planets,
  Stars, and Stellar Systems'', Vol 6, 2011.}  

The dust-corrected $B/T$ histograms for the morphological type bins
are shown in Figure 8: though finding an overlap in $B/T$ among the
Hubble types is not new, this figure clearly demonstrates how $B/T$ in
S0s extends to the region of the late-type spirals, thus covering the
full $B/T$ range to near zero values. To our knowledge this is
  the first observational evidence in support of the scenario
  suggested by van den Bergh [72], according to which the S0s might form a
  sequence of S0a, S0b and S0c galaxies in a similar manner as
  spirals.

\subsection{Low B/T in conjunction with semi-analytic models (e.g. Khochfar et al. [41])}

Formation of early-type galaxies (ETC) has been recently modeled by
Khochfar et al. [41] by semi-analytical models (SAM), taking into
account the new 3DAtlas kinematic observations by Emsellem et
al. [29], defining the fast and slow rotators. In their models $B/T$
is a critical boundary condition to define the ETCs. The range 0.5 $<$
$B/T$ $<$ 0.9 is used for 80$\%$ of the fast rotators, and 20$\%$ of
them have $B/T<$0.5, for which galaxies also the gas fraction was used
as a selection criterion. The galaxies with low $B/T$ are similar to
late-type spirals, called 'red spirals', except that they have barely
no gas, and therefore have no spiral arms. In their models the
majority of fast rotators have accreted their bulge mass in minor
mergers, whereas the 'red spirals' were formed by starvation in more
dense galaxy environments.

The above models seem to explain well the distribution of
$\lambda_{Re}$, a proxy of specific baryonic angular momentum content
in galaxies.  The difference between fast and slow rotators is largely
based on the stellar disc fractions in galaxies.  This is consistent
with our morphological analysis of NIRS0S in which, besides S0s, also
most of the galaxies classified as late-type ellipticals in RC3 show
disk structures.  However, in many S0s also small central components
appear in the regions covered by the kinematic observations. This
complicates the fast/slow interpretation, because nuclear bars, lenses
and rings might have different velocity dispersions. The richness of
the central structures in S0s might partly explain why $B/T$ does not
correlate with $\lambda_{Re}$, which is shown in Figure 9 (above), 
  e.g., galaxies with larger $B/T$ do not have smaller
  $\lambda_{Re}$. Also, there is no correlation between
$\lambda_{Re}$ and S\'ersic index $n$, as shown in Figure 9 (below),
e.g.  galaxies with disk-like pseudo-bulges (with small $n$) do
  not have a tendency to be more rotationally supported.

Good news in the models by Khochfar et al. [41] is also that it
predicts a fairly large range of $B/T$s for the fast rotators. However, the
used $B/T$ range of 0.5-0.9 is almost completely outside the range
obtained by the multi-component decompositions for the S0s
(see Fig. 8). In the literature possible mechanisms have been discussed
which might lower the $B/T$ in SAMS: (a) increasing the energy
injected by supernovae ([2]), or (b) taking into account mass
loss from already existing stars, which can lower $B/T$ by a factor of
2-3 ([56]). However, these explanations have been
criticized by Scannapieco et al. [65]. The first because during the
mass loss the bulge density is also reduced, and the density of the
disk increases only in the central parts of the disk. And the second,
because the suggested efficiency of SN feedback is too high compared
to the cosmological models, e.g., by Guo et al. [61]. In the current
models by Khochfar et al. lower $B/T$ values for fast rotators are obtained by
significantly decreasing the frequency of minor mergers, but that might not be
consistent with the observed merger frequency in the Universe.

\subsection{Low B/T in conjunction with tidal models (e.g. Bekki $\&$ Couch [11]) }

In the stripping models the $B/T$ problem has generally been the
opposite: it has been difficult to make such massive bulges as
observed in galaxies.  In Bekki $\&$ Couch [11] the problem was
solved by assuming that the tidal effects increase the efficiency of
star formation and might also add some fresh material into the
galaxies during the encounter process. Here we discuss how well
the $B/T$ ratios obtained in NIRS0S are compatible with these models.

Compared to Khochfar et al. [41], in Bekki $\&$ Couch [11] a lower
$B/T$ range of 0.3-0.5 was used for the S0s. This is well within the
range of the measured $B/T$ values in Figure 8, but still lacking the
lower $B/T$ tail.  In the simulations by Bekki $\&$ Couch, $B/T$ is
measured kinematically from the simulations by linking each simulated
star either to a bulge or a disk. Based on the comparison by
Scannapieco et al. [65], such kinematic decompositions give
significantly higher $B/T$ values than the decomposition of the flux
distribution in galaxies, done in a similar manner as in
observations. A similar conclusion was made also by Laurikainen $\&$
Salo [85].  Taking this into account the $B/T$ range in the models by
Bekki $\&$ Couch might be even more closer with the observed $B/T$
range.

\section{Properties of bulges and disks, based on the decompositions}

\subsection{Properties of bulges and disks: S0s vs. spirals}

The photometric properties of bulges and disks were compared between
S0s and spirals by Laurikainen et al. [53], without including any
kinematic information. The parameters were corrected for the internal
dust in galaxies. The final sample, with reliable decompositions
made, consists of 340 galaxies.  Naively, strong correlations between
the parameters of the bulge and the disk are expected if the disks
were formed first, and the bulges emerged from the disc by secular
evolution. On the other hand, in hierarchical clustering, followed by
galaxy mergers, the properties of bulges were established already
during the last major merger event ([39]), and presumably changed very
little after that.  Also, if the formative processes of bulges in S0s
were similar to those of the elliptical galaxies, a continuity in the
properties of bulges and ellipticals is expected.

It was found that the bulges of S0s are different from the elliptical
galaxies, and similar to the bulges in spirals, manifested in the
$r_{eff}$ (bulge) vs. $M_{bulge}$ diagram (Fig. 10): bulges both for
the S0s and spirals appear below the line for the elliptical galaxies,
indicated by the dashed line\footnote{the dashed line is not for the
  Coma cluster ellipticals, as erroneously claimed in Laurikainen et
  al. [53]}. It was shown by Graham $\&$ Worley ([35], their
Fig. 11) that this line nicely follows the observed distribution of
the ellipticals (see [34]), covering a large magnitude range, and
appearing in various types of environments.  For spirals the
discontinuity of bulges and ellipticals was shown also by Gadotti
[33].  The fact that the bulges of S0s are not similar to the
elliptical galaxies is manifested also in the Kormendy relation and in
the photometric plane where only the brightest bulges of S0s follow
the location of the Coma cluster ellipticals (see Figs. 2 and 3 in
[53]).

The $r_{eff}$ (bulge) vs. $M_{bulge}$ in Figure 10 extends to the
regime of very small bulges, corresponding to the effective radii of
the dwarf early-type galaxies, dEs.  In Figure 10 the dEs would appear
in the flat part of the dashed line (see [34]). However,
it is worth noticing that the small bulges in S0s have several
magnitudes higher surface brightnesses than dEs, and therefore do not
fit to the scenario in which dEs were formed from Sc type spirals by
loosing their disks in galaxy harassment in dense cluster
potentials. This is consistent with the recent study by Kormendy
  $\&$ Bender [88], who suggested that the red and dead dwarf
  galaxies are rather transferred from later type Scd-Im galaxies.

Not only bulges, but also the disks in S0s are similar to those in
spirals. It was shown by Laurikainen et al. [53], that both for the
S0s and spirals the scale length of the disk ($h_R$) varies between
1-10 kpc, and the central surface brightness ($\mu_0$) between 16-20
magnitudes.  Also, $\mu_0$ has no correlation with the mass of the
disk ($M_{disk}$). It was concluded that $\mu_0$ does not follow the
Freeman law ([32]) pertaining a constant central surface brightness
(see Fig. 9 in [53]).  The similarity in the disc sizes between S0s and
spirals is consistent with the tidal models by Bekki $\&$ Couch [11]
for the origin of S0s.  They stated that ``it is unlikely that S0s
have systematically larger disk masses in comparison with spirals.''

For the S0s the photometric properties of bulges and disks in NIRS0S are found
to be coupled, in a similar manner as for spirals [53]. This is
manifested in correlations between their scale parameters, $h_R$ vs.
$r_{eff}$(bulge), and their masses, $M_{bulge}$ vs. $M_{disk}$, as shown in
Figure 11. Also, for the bulges having
$M_{bulge}$ = -20 mag in the $K_s$-band the difference in $M_{disk}$
between the early (S0, S0/a, Sa) and late type (Sab and later) galaxies is
nearly one magnitude. Based on the stellar population models by
Bedregal, Arag\'on-Salamanca $\&$ Merrifield  [10] this difference could be explained by quiescent
star formation if star formation of the disc in spiral galaxies was
truncated 1-6 Gyr ago.

These results are consistent with the picture in which the discs were
formed first and the bulges emerged from the disc material afterwords,
without invoking any merger events, which might destroy the
observed correlations between the bulges and disks. However, our
results do not unambiguously rule out the other scenarios. For example,
in minor mergers the satellite galaxies increase the bulge mass in one hand, and
at the same time the disks stars are spilled to a larger radius, which
might also lead to a correlation between the scale parameters of the bulge
and the disk.

\subsection{Properties of bulges in the Hubble sequence}
 
Figure 12 shows the mean total galaxy brightnesses in the $K_s$-band and
the parameters of the bulge in the Hubble sequence [53].  The bulge
parameters are compared with those by Graham $\&$ Worley [35]. Notice
that the comparison is meaningful only between T=0-5 where enough
common galaxies appear in both studies.

The following things can be noticed: (1) in the near-IR, tracing the
old stellar population of galaxies, the S0s in NIRS0S are not less
luminous than the spirals in OSUBSGS (both samples are
magnitude-limited, selected in a similar manner). A similar result in
the near-IR was obtained also by Burnstein et al. ([89]). This is a
puzzle if S0s were formed from spirals Sc or earlier, simply by
loosing their gas in the stripping mechanisms, followed by quenching
of star formation. The luminosities of S0s could be explained if they
have gained more mass  after the S0 characteristics were
settled down, for example by accreting a significant number of 
satellite galaxies during their lifetime. However, that is not supported by our
analysis of the $Q_g$ distribution of bars in S0s, at least if the
accreting satellites were gas rich. A more likely solution to this
puzzle is that at higher redshifts some fraction of the progenitors of
S0s were more luminous than the local spirals. This is evidenced by
Geach et al. [90] showing that LIRGS at z=0.5 could account for the
star formation needed to explain the bright end of the S0 galaxy
luminosity function (see the discussion in [53]).

(2) $r_{eff}/h_R$ decreases slightly from S0s towards spirals.  This
parameter is smaller than obtained in the previous studies, mainly
because we use multi-component decompositions which presumes the
erroneous additions of bars to the bulges.  For all Hubble types the
observed mean $r_{eff}/h_R$ is smaller than predicted by the current
merger models (e.g.  [58]). (3) The S\'ersic index $n$ is very similar
for S0, S0/a and Sa galaxies, and slightly decreases towards the later
types. Such small values of S\'ersic index ($n$ $\le$ 2) are not
predicted in the current merger models (e.g.  [66], [59]). (4) $B/T$
was already discussed in Section 5. Figure 12 additionally shows that
the mean $B/T$ is very similar for S0, S0/a and Sa galaxies, and only
after that decreases towards the later types. It was also shown by
Laurikainen et al. [51] that $B/T$ is lower for barred than non-barred
galaxies, and also lower for galaxies having flat-top DG bars compared to
the more simple SG bars.

Overall, the above properties of bulges are not well compatible with
the current merger models.

\section{The outlined picture}

\subsection{S0s transformed from spirals: the Hubble sequence revisited}

The Hubble sequence is revisited, following the early suggestion
  by Spitzer $\&$ Baade [71] and van den Bergh [72]: namely the S0s
  are suggested to be spread throughout the Hubble sequence in
  parallel tuning forks as spirals (S0a, S0b, S0c etc.)([53],
  [54]). Originally this scenario was arisen by van den Bergh because
  he found that the entire S0 class was offset towards fainter optical
  magnitudes from the luminosity distribution of Sa galaxies. He then
  assumed that there must exist anemic S0s whose surface brightness
  distributions are similar to the faint surface brightness levels of
  Sb and Sc galaxies. However, at that time no such galaxies were
  found. Our NIRS0S/OSUBSGS analysis has shown that indeed, the disks
  in S0s have a similar range of central surface brightnesses and
  scale lengths of the disk as the disks of Sa-Sc type spirals. And in
  particular, our multi-component decomposition approach has shown
  that the bulges in many S0s have $B/T$ ratios as small as typically
  found in Sc type spirals [53]. A similar view to the Hubble sequence was
  taken recently also by Cappellari et al. [22], based on new
  kinematic observations, showing that bulges in S0s are fast rotating
  oblate systems. During the referee process of this work a paper by
  Kormendy $\&$ Bender [88] appeared in astro-ph, in which a similar
  view was also suggested.

Our results are consistent with the idea according to which S0s are
mainly former spirals transformed into S0s by loosing their gas in
some stripping mechanisms. 
The following results of our NIRS0S/OSUBSGS
analysis are consistent with this picture: (a) S0s cover the full
observed $B/T$-range of spirals.  (b) The photometric properties of
bulges and disks appeared to be similar in S0s and spirals. And most
importantly, the bulges in S0s have no continuity with the elliptical
galaxies in the $r_{eff}$ (bulge) vs. $M_{bulge}$ diagram. (c) The
relative masses and scale parameters of the bulges and disks in S0s
are coupled, in a similar manner as in spirals, which hint to a common
origin of the bulges and disks in these galaxies. What needs to be explained
in this scenario is why the S0s in NIRS0S are not less luminous than their assumed spiral
progenitors.

\subsection{Solution for the bulge problem in the stripping scenario?}

In the stripping scenario, a problem of larger bulges in S0s, compared
to bulges in spirals, was discussed. It was concluded, based on the
stellar population models by Bedregal, Arag\'on-Salamanca $\&$
Merrifield [10], that spirals with bulges brighter than -20 magnitudes
in the $K$-band are massive enough to be converted into bulges of S0s,
simply by slow internal galaxy evolution.  However, that cannot
explain why S0s are not dimmer than the spirals, as expected if they
lost some of their gas in the stripping process. This implies that
some other mechanism is also needed to complete the mass accretion in
the bulges of S0s. Possible alternatives are that the progenitors of
S0s were more gas rich than the present day spirals, which would
  explain also the similar total galaxy brightnesses in most S0s,
  compared to the galaxy brightnesses in spirals, in the near-IR. Or
some mechanism other than gas accretion is taking place. One
suggested mechanism is the so called potential-density phase shift
mechanism suggested by Zhang [74], allowing stellar mass transfer even
after the gas depletion has stopped. Or, as suggested by Bekki $\&$
Couch [11], tidal effects in galaxy groups can increase star formation
efficiency in the central regions. Minor mergers can also accret
material to the main galaxies, but that process needs to be mild
enough for not to increase the S\'ersic index over the fairly low
observed values, at least in the current models. Also, if our
conclusion of the $Q_g$ distribution for the S0s is correct, the
accreting satellites need to be fairly gas poor.

\subsection{Bars and lenses: evidence of secular evolution? }

A test was made for the simulation models by Bournaud $\&$ Combes
[14], in which bars are expected to have multiple episodes during the
Hubble time.  Based on the observed bar strength ($Q_g$) distribution
we concluded that the S0s have not accreted gas for a long
time. Spirals show a larger number of strong bars, but the observed
$Q_g$ distribution even for them resembles the non-accretion
models. This implies that most probably bars are fairly robust in all
Hubble types, without experiencing multiple bar episodes in the Hubble
time. However, inspite of that bars may evolve significantly over
time, as manifested in the morphology of the old stellar population of
bars in the near-IR. Namely bars in S0s have ansae morphology and
double-peaked Fourier density profiles much more frequently than bars
in spiral galaxies.  According to the dynamical models such features
can be associated to evolved strong bars. Bars may also dissolve into
lenses, as first suggested by Kormendy [43], which is consistent with
the observed lower bar fraction, and a larger fraction of lenses among
the S0s. However, based on the morphology of lenses, not all lenses
can be dissolved former bars.

A new lens type, barlens, was introduced, appearing in strong bars,
most probably forming part of the bar itself. We assume that it might
be related to the re-distribution of matter in galaxies.  Also,
X-shaped bars in nearly face-on galaxies were detected, which cannot
be simply vertically thickened bars by buckling effects, and therefore
needs to be explained in the framework of the theoretical
models. Multiple lenses were detected even in 25$\%$ of the S0s,
including barred and non-barred galaxies, which are not yet well
understood in the current paradigm of galaxy formation and
evolution. We anticipate that secular evolution might play an
important role in the formative processes of lenses.
  
\clearpage
\newpage

\begin{figure}

\includegraphics[width=160mm]{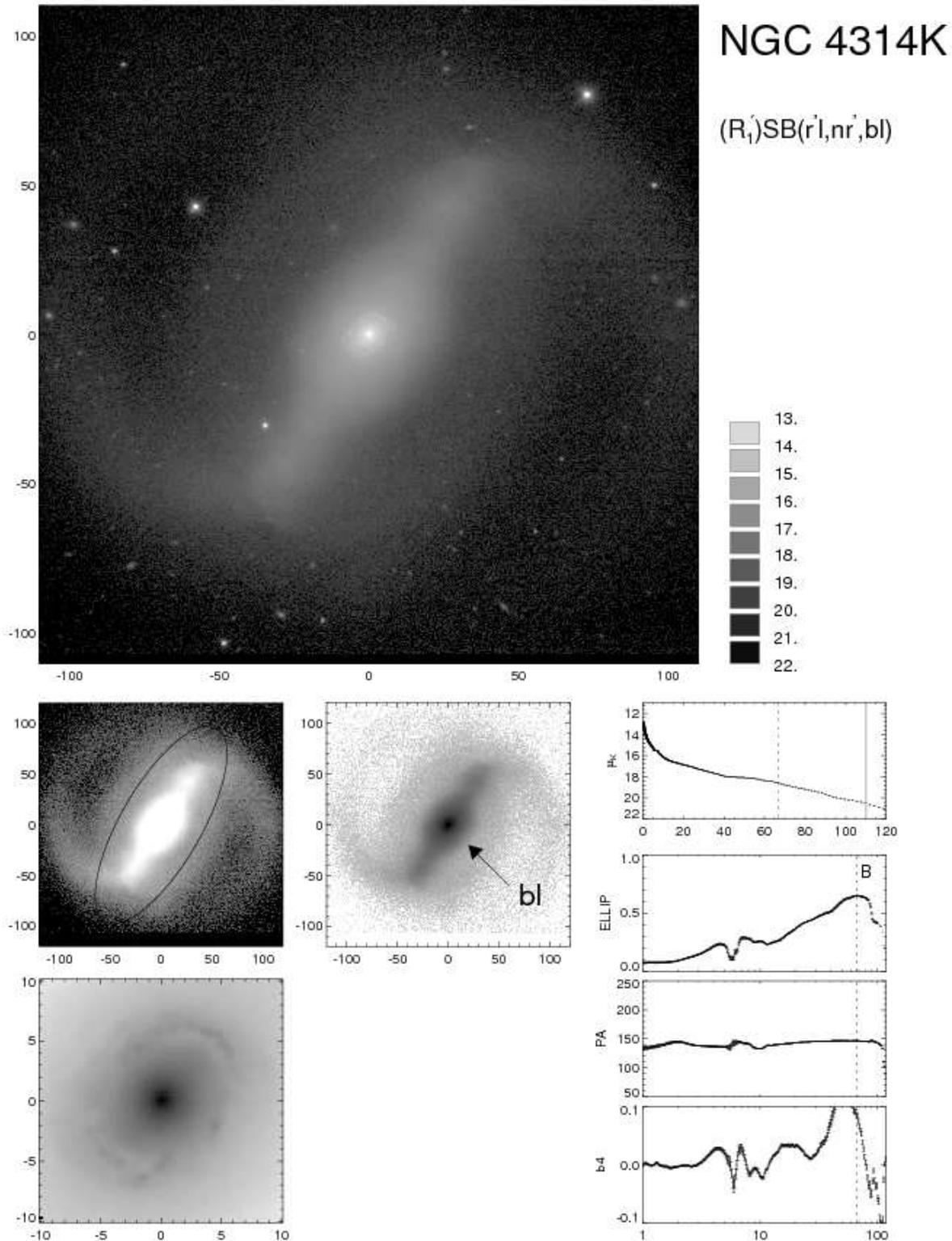}
\caption{An example of the Atlas image [54]. The upper panel shows the
    whole galaxy using the full magnitude range, for which the scale
    is indicated in the right. The scales in x- and y-axis are in
    arcseconds.  The small panels show the same image, but selecting
    the radial and magnitude scales in a different manner:
    the two upper panels use the full radial scale, whereas in the
    lower panel only the central region of the galaxy is shown. The
    morphological classification is from the Atlas paper. The arrow
    indicates the barlens.  The graphics show the radial profiles
    of the surface brightness (upper panel), ellipticity (second 
    panel), position angle (third upper panel), and parameter b4
    (lowermost panel) indicating deviations from perfect ellipticity. The
    logarithmic radial scale is in arcseconds. In these plots the
    vertical dashed lines indicate the bar radius, and the full line
    is the radius where the surface brightness is 20 mag
    arcseconds$^{-2}$.  Originally the figure appeared in Laurikainen
    et al. [54].}
\label{figure-atlas}
\end{figure}
\clearpage
\newpage

\begin{figure}
\includegraphics[width=170mm]{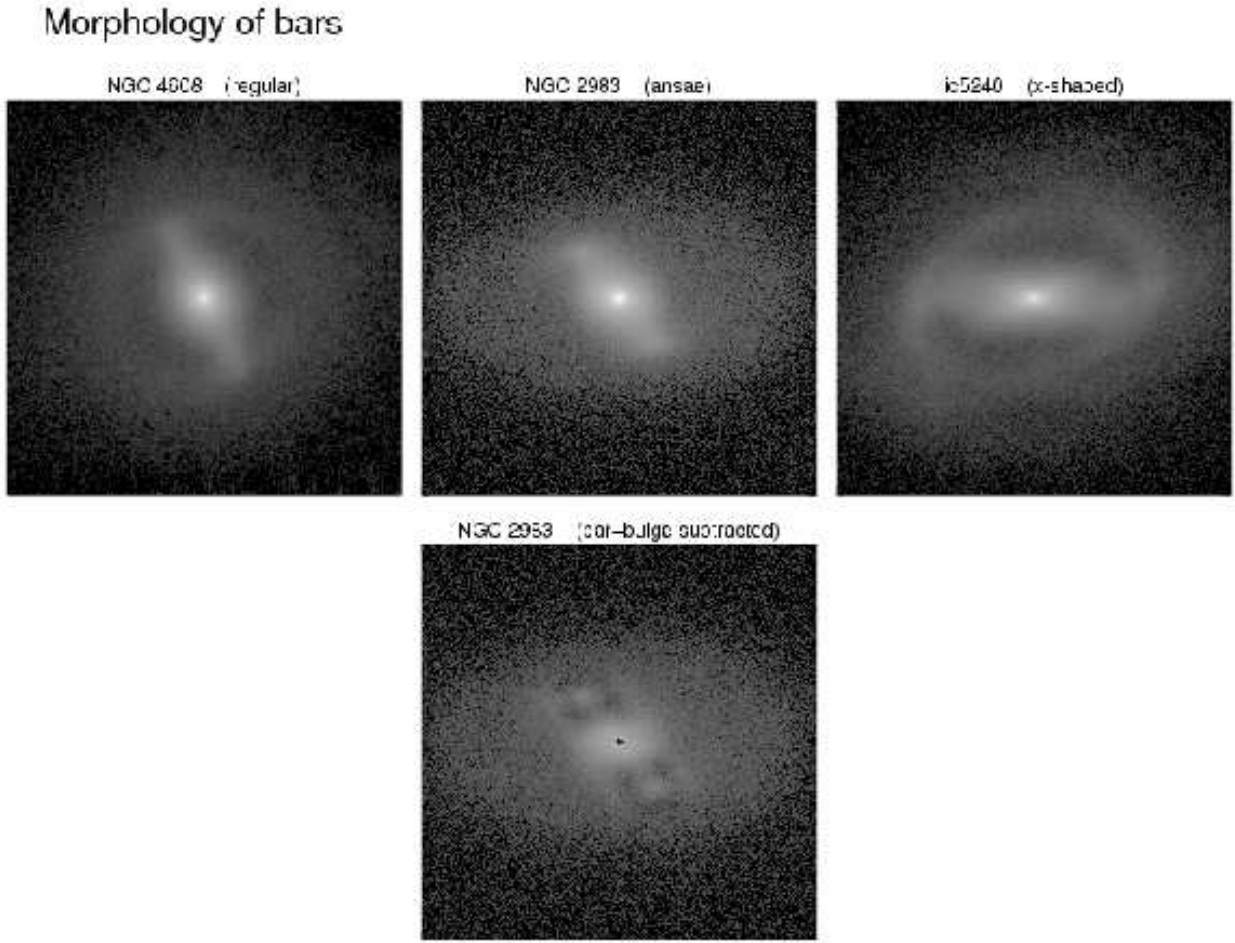}
\caption{This figure is compiled of two figures that originally appeared
in Laurikainen et al. [54]. The images are flux calibrated and are shown in a
logarithmic scale, covering the full magnitude range of the images. 
In the upper row the original images are shown, whereas in the lower
row the bar and bulge models from the decomposition are 
subtracted from the original image of NGC 2983. These images highlight the main
type of bar morphologies in S0 galaxies.}
\label{figure-barmorphology}
\end{figure}
\clearpage
\newpage

\begin{figure}
\includegraphics[width=170mm]{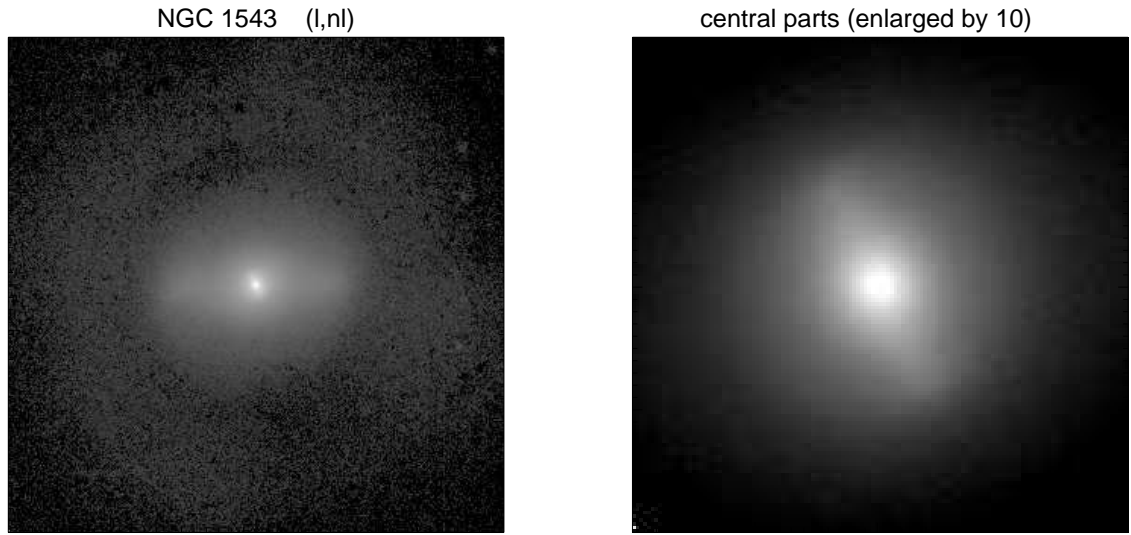}
\caption{Two panels from Figure 15 in Laurikainen et al. [54] are
shown.  In the {\it left panel} the whole galaxy is shown using the full
magnitude range of the image. In this image the outer lens and the
inner lens surrounding the main bar are clearly visible.  In the
{\it right panel} only the central regions of the galaxy is shown
(enlarged by a factor of 10),
demonstrating the morphology of the nuclear bar, and the nuclear lens
extending to the radius of the nuclear bar.}
\label{figure-doublebar}
\end{figure}
\clearpage
\newpage

\begin{figure}
\includegraphics[width=170mm]{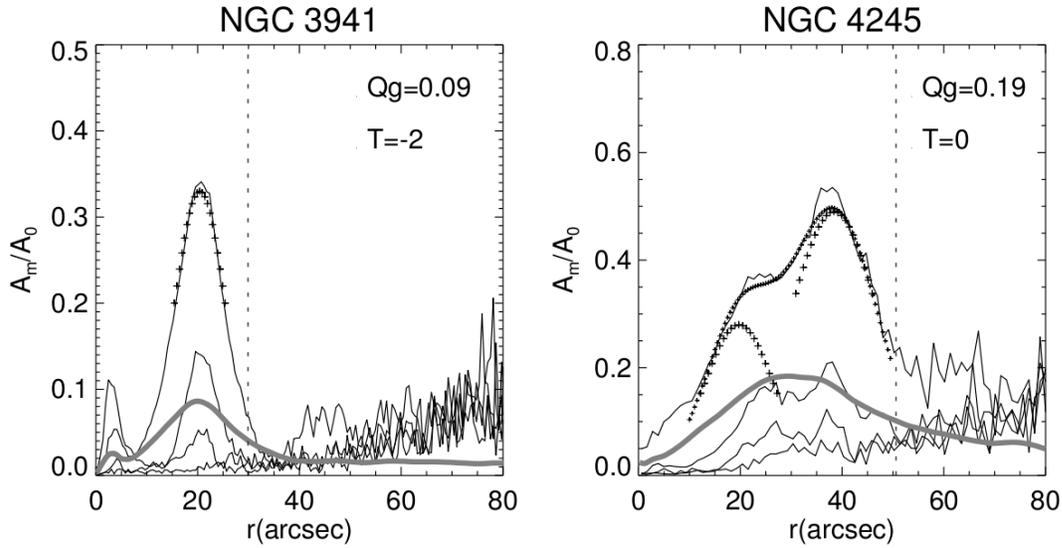}
\caption{Radial profiles of the Fourier amplitudes of density: {\it left
panel} shows a simple bar, whereas {\it right panel} shows a
two-component bar: both lower and higher Fourier modes are significant
in these two components.  The thin lines show the $m$=2, 4, 6, 8 Fourier
amplitudes, normalized to $m$=0. The thick grey line shows the force
ratio $QT(r) = |FT(r,\phi)|max / <|FR(r,\phi)|>$ ($Q_g$ indicates the
maximum of QT ), and the vertical line indicates the length of the
bar. The crosses show individual Gaussian fits to the $m$=2 Fourier
amplitude profile. The
single peaked bar represents a thin classical bar. The double peaked
amplitude profile is associated to a bar having a shorter and
vertically thicker inner part, and a longer and vertically thinner
outer part. This figure is a modified version of Figure 8 in Laurikainen et al. [51]. }
\label{figure-gaussians}
\end{figure}
\clearpage
\newpage

\begin{figure}
\includegraphics[width=170mm]{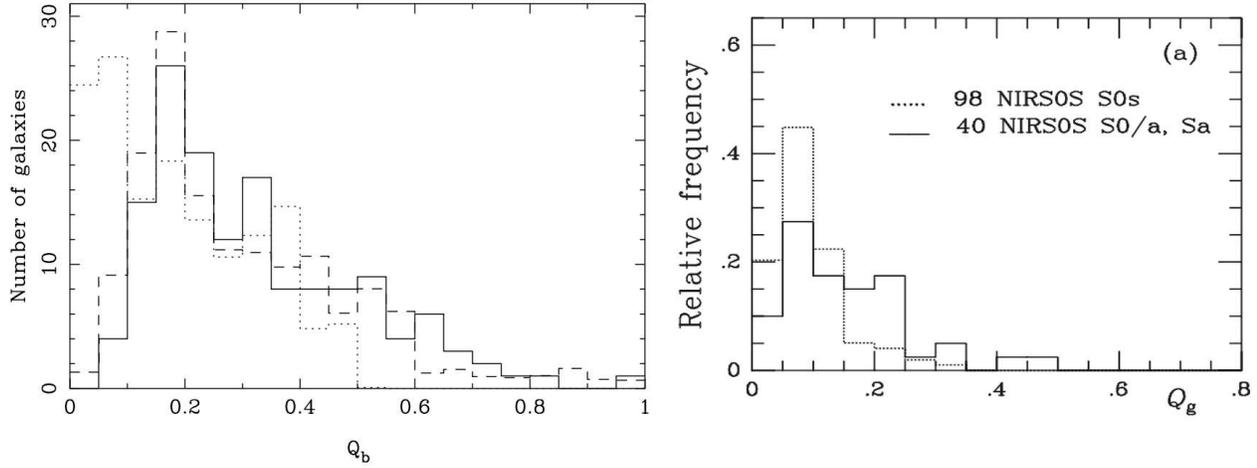}
\caption{ {\it Left}: the observed $Q_g$ distribution for spirals
    in the OSUBSGS (cover the Hubble types up to T=9), taken from Block
    et al. [80], compared with the simulation models by Bournaud $\&$
    Combes [14]. The solid line shows the observed values from Block
    et al., the dotted line is the model with no gas accretion, and
    the dashed line shows the model that doubles the total galaxy mass
    in 10$^{10}$ years.  {\it Right}: the observed $Q_g$ distributions
    for the S0s and early-type spirals in NIRS0S, taken from Buta et
    al. [86].  Notice that $Q_b$ in Block et al. [80] in the left
    panel has the same meaning as $Q_g$ in Buta et al. [86] in the
    right panel.}
\label{figure-Qgdistributions}
\end{figure}
\clearpage
\newpage

\begin{figure}
\includegraphics[width=170mm]{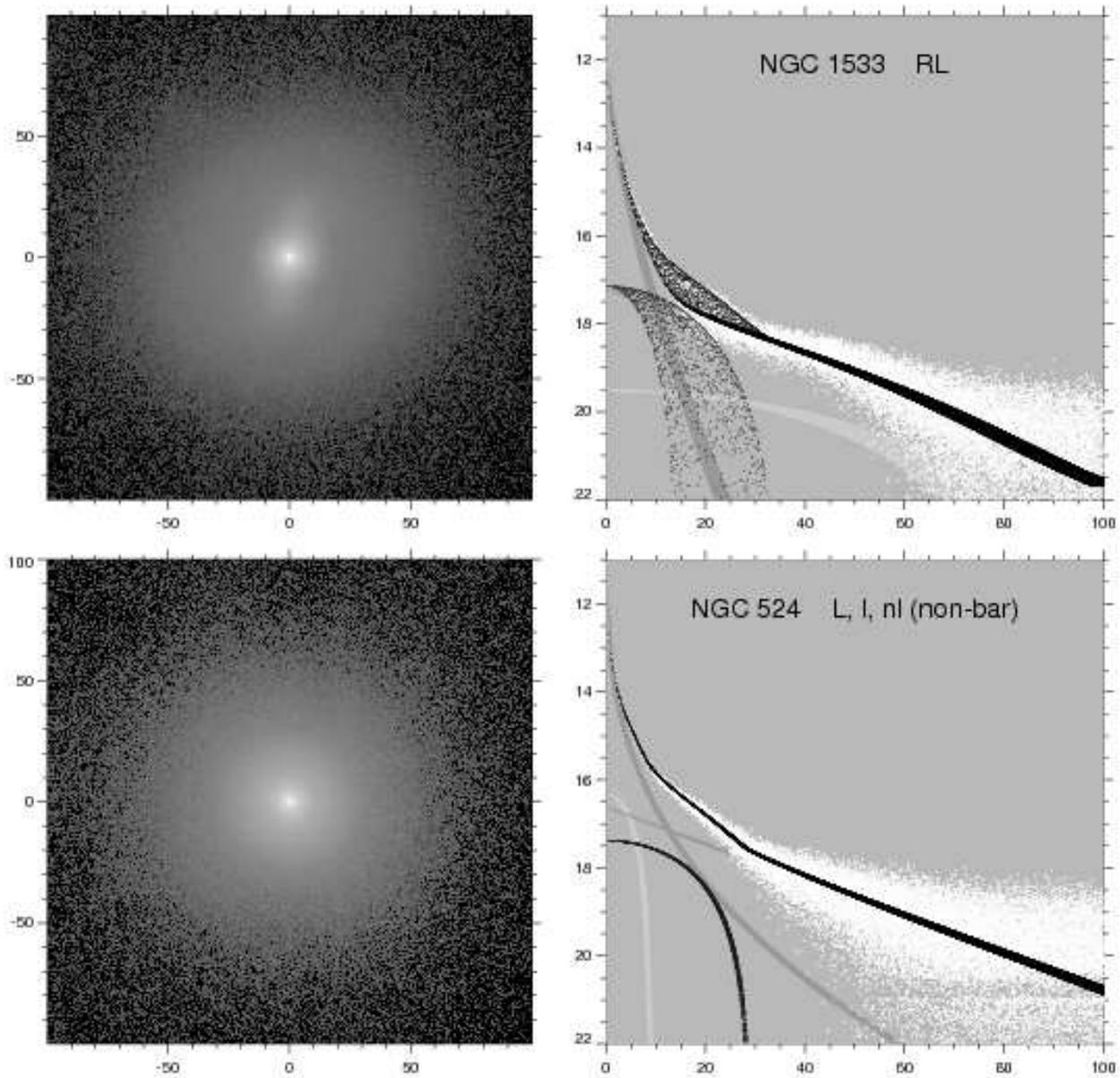}
\caption{Examples of non-barred lens galaxies. {\it Left panels} show
  the images, whereas {\it right panels} show the 2-dimensional surface
  brightness profiles, in which over-plotted are the fitted functions
  to the structural decompositions, as well as the total model. White
  dots are for the image pixel values, and the black lines/dots are for
  the total model. Models for the bulges are shown by dark grey lines,
  and the models for the lenses by light grey and black dots in the
  inner regions.  For example, in NGC 524 the two curved inner
  components are the nuclear lens (nl) and lens (l), and the outer
  exponential function fits the outer lens (L). This figure was
  originally published by Laurikainen et al. [54]. }
\label{figure-lenses}
\end{figure}
\clearpage
\newpage

\begin{figure}
\includegraphics[width=170mm]{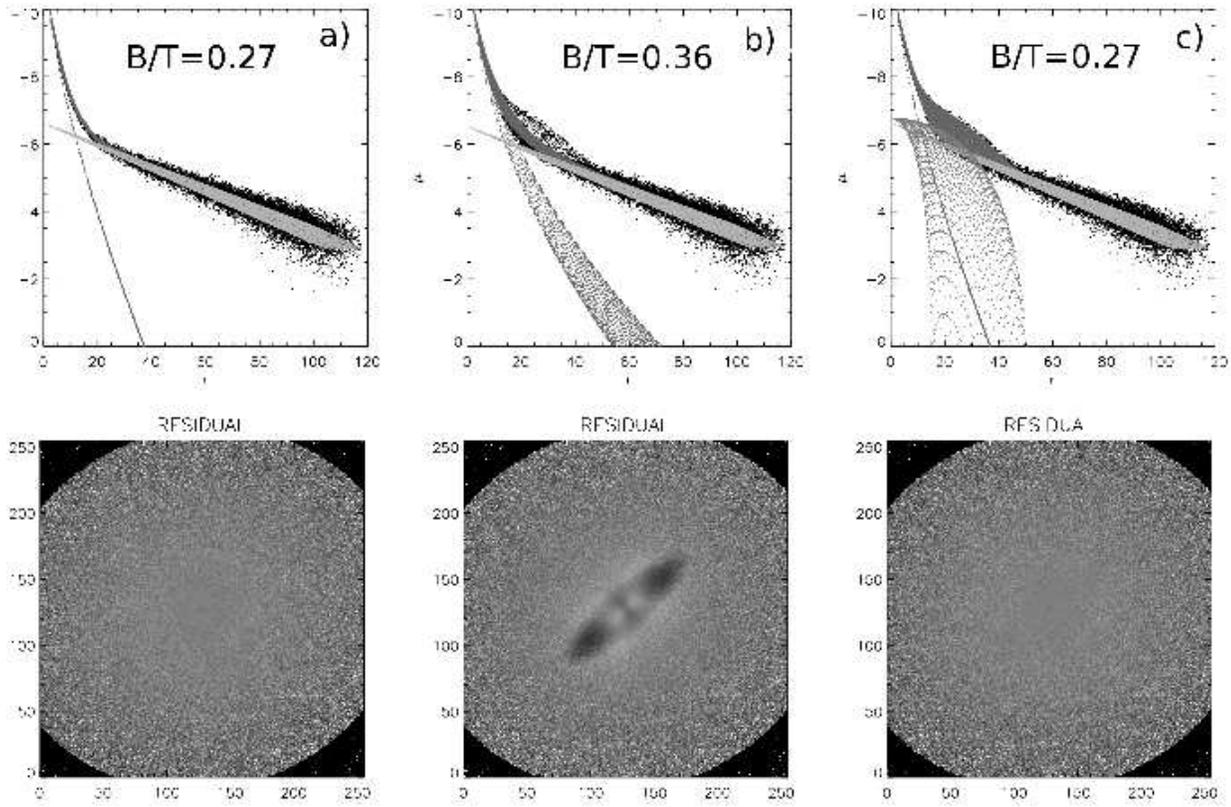}
\caption{{\it Upper row}: surface brightness profiles of synthetic
  images in which noise is added to mimic a real image. In a) the
  image has only a bulge and a disk, whereas in b) and c) a small bar
  is added on top of the disk. Black dots show the pixel values of the
  synthetic images, and the other symbols show the functions fitted to
  the 2-dimensional images, as explained in the text. {\it Lower row}:
  the residual images for the above decompositions are shown.  They
  are obtained by subtracting the total decomposition model from the
  original synthetic image.  The figure is a modified version of
  Figure 2 in Laurikainen et al. [49]. }
\label{figure-test1}
\end{figure}
\clearpage
\newpage

\begin{figure}
\includegraphics[width=170mm]{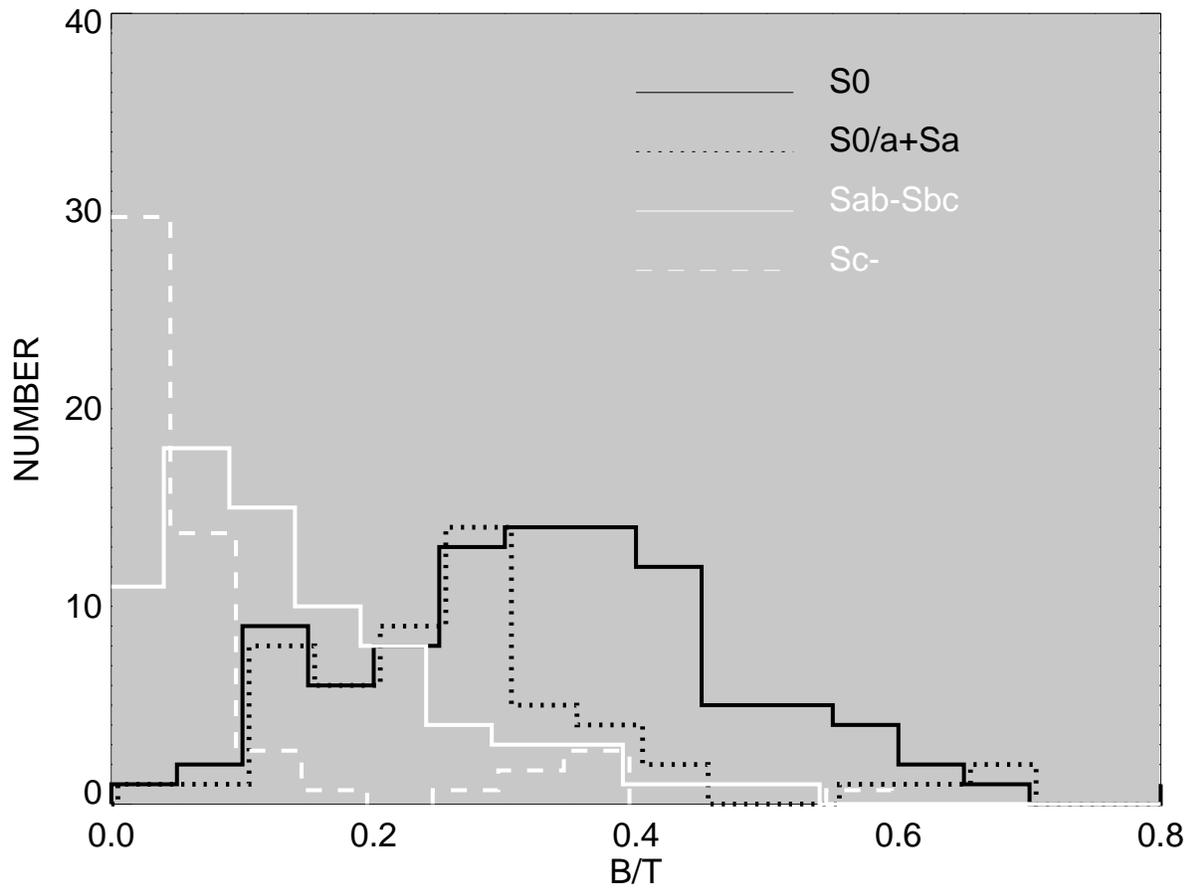}
\caption{Histogram of the $B/T$ flux ratios in $K_s$-band, corrected
  for Galactic and internal dust extinction. The corrections are different for
  the bulge and the disk. The histograms for the S0s (full line), and
  the S0/a galaxies (dotted line) are based on the NIRS0S sample. For
  spirals the lines for Sab-Sbc (full white line) and later than Sc
  (dashed white line) are drawn using the OSUBSGS sample. This figure
  was originally published by Laurikainen et al. [53]. }
\label{figure-histogram}
\end{figure}
\clearpage
\newpage

\begin{figure}
\includegraphics[width=170mm]{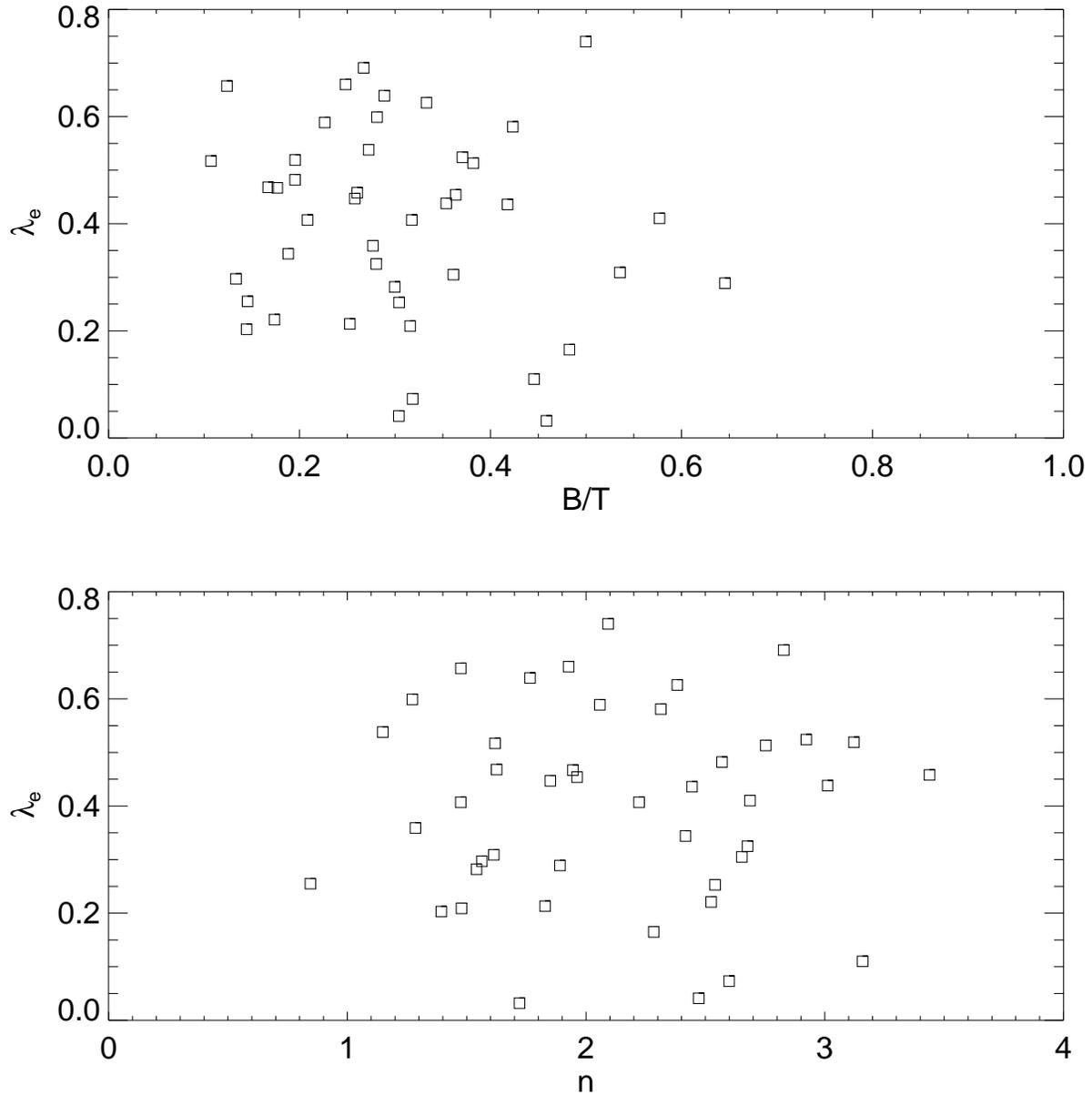}
\caption{The kinematic parameter $\lambda_R$ vs. $B/T$ ({\it upper panel}),
  and vs. S\'ersic $n$ ({\it lower panel}). Parameter $\lambda_{Re}$, taken
  from Emsellem et al. [29] can be used as a discriminator
    between oblate and triaxial nature of the galaxies, within the
    apertures used in the measurements. In this plot $\lambda_R$
    obtained within one effective radius of the galaxy is used. The
  $n$-parameters and the dust corrected $B/T$ values are taken from
  Laurikainen et al. [53].}
\label{figure-kinBTn}
\end{figure}
\clearpage
\newpage

\begin{figure}
\includegraphics[width=170mm]{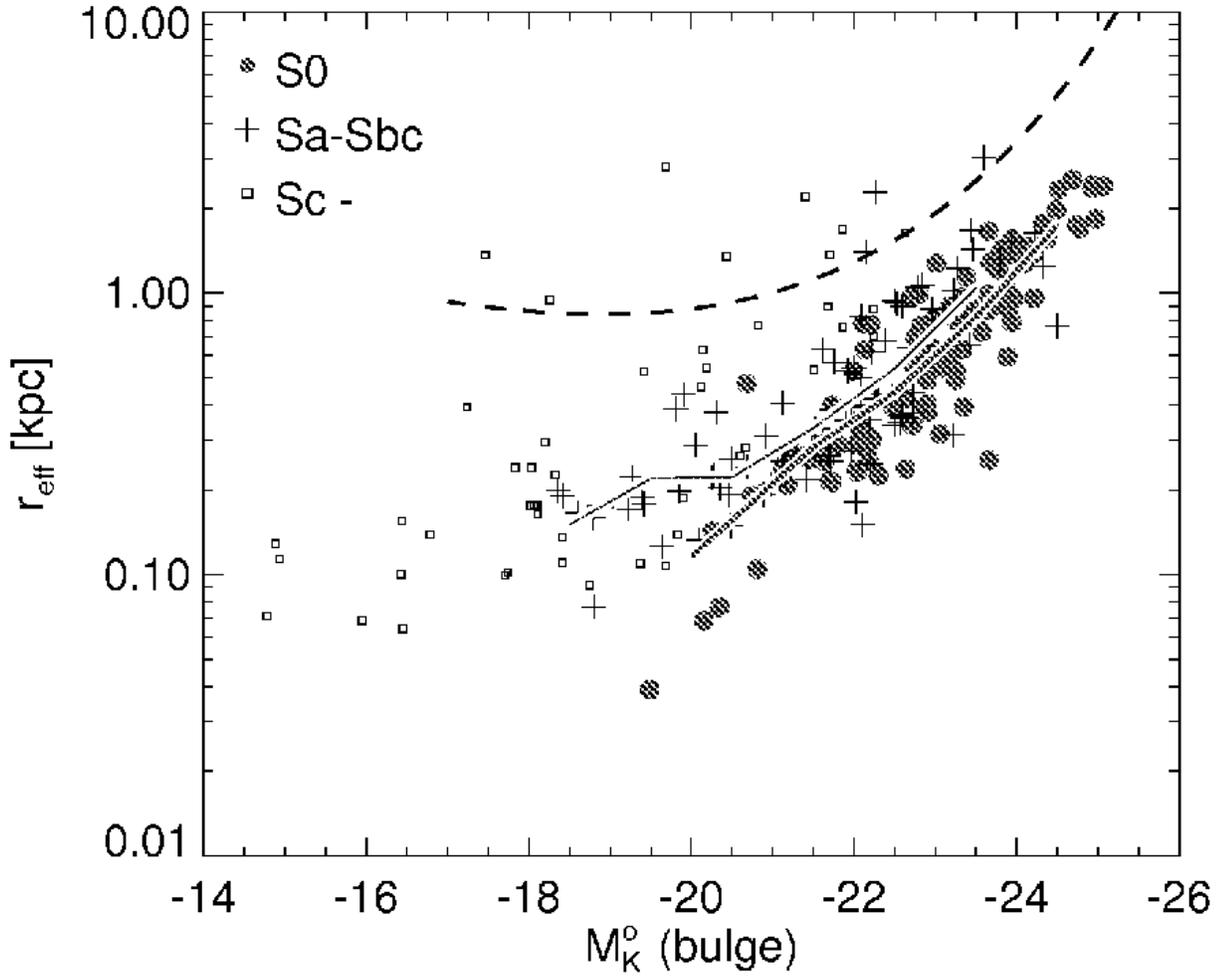}
\caption{ The effective radius of the bulge ($r_{eff}$) plotted
  against the absolute brightness of the bulge ($M_K^o(bulge)$). Shown
  separately are: S0s (filled symbols), Sa-Sbc spirals (crosses),
  and Sc and later types (squares).  The dashed line shows the
  location of the elliptical galaxies, taken from Graham $\&$ Worley
  (2008). The continuous lines show the mean values of the data points
  in one magnitude bins: the thick line is for the S0s and the
  thin line for the Sab-Sbc spirals. This figure is part of Figure 8a
  in Laurikainen et al. [53]. }
\label{figure-reffMbulge}
\end{figure}
\clearpage
\newpage

\begin{figure}
\includegraphics[width=170mm]{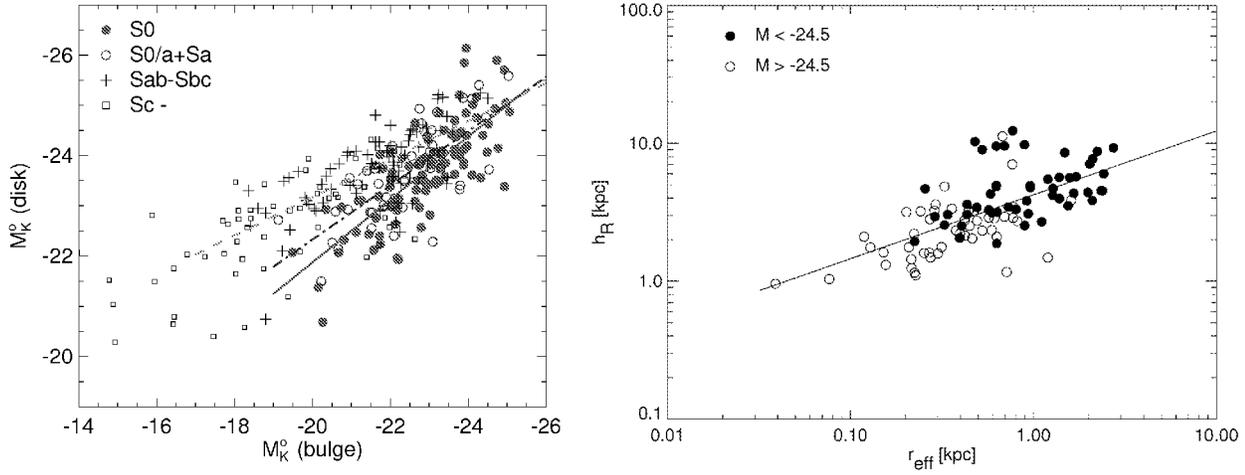}
\caption{ {\it Left}: the absolute brightness of the disk
  ($M_K^o(disk)$) is plotted against the absolute brightness of the
  bulge ($M_K^o(bulge)$). (a) S0s (red filled circles), S0/a+Sa (open
  circles), Sab-Sbc (crosses) and Sc and later types (squares). Solid,
  dash-dotted and dashed lines show linear fits for the S0s, S0/a+Sa
  and Sab-Sbc galaxies, respectively. All the correlations are
  statistically significant. {\it Right:} $h_R$ (disk) vs. $r_{eff}$
  (bulge) for the galaxies in NIRS0S, divided to two galaxy brightness
  bins in $K_s$-band. This figure is compiled from two figures
  originally published by Laurikainen et al. [52], [53]. }
\label{figure-scaling} 
\end{figure}
\clearpage
\newpage

\begin{figure}
\includegraphics[width=170mm]{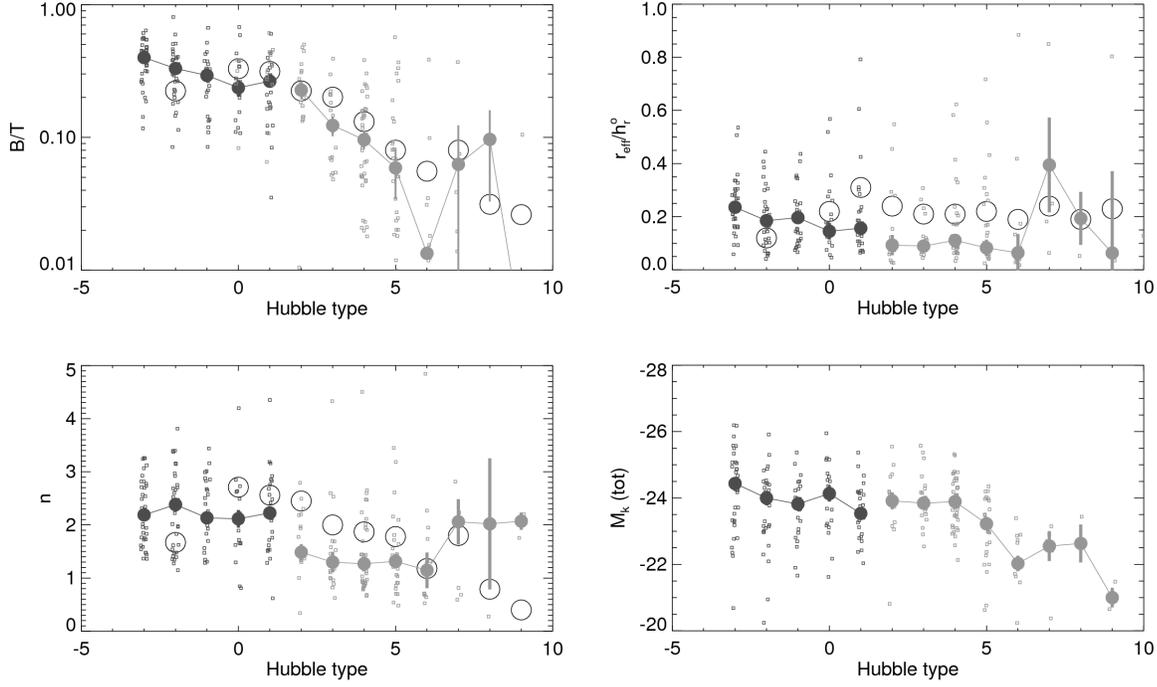}
\caption{ As a function of Hubble type are shown: (a) the dust
  corrected $B/T$, (b) S\'ersic index $n$, (c) the ratio of the
  effective radius of the bulge ($r_{eff}$) divided by the scale
  length of the disk ($h_r^o$), and (d) the total absolute galaxy
  magnitude corrected for Galactic extinction ($M_{tot}^o$). The {\it
    small symbols} show the measurements for the individual galaxies
  (with a small random horizontal offset), {\it filled circles} show
  the median values in each Hubble type bin, and {\it large open
    circles} are the median values from Graham $\&$ Worley ([35];
  their value for S0s refers to types -3, -2, -1, and is here drawn at
  T=-2).  The error bars are taken to be the sample variances in each
  bin divided by square root of the number of galaxies in each bin.
  Dark symbols are for the galaxies in NIRS0S and light symbols for
  the galaxies in OSUBSGS. In a slightly different form the same
  figure was originally published by Laurikainen et al. [53].  }
\label{figure-Hubblesarja} 
\end{figure}
\clearpage
\newpage

 [1] Afanasiev V.L., Silchenko O.K. 2007, Astronomical $\&$ Astrophysical Transactions, 26, 4, 311-337

 [2] Agertz O, Teyssier R., Moore B. 2011, MNRAS, 410, 1391

 [3] Aguerri J.A.L., Balcells M., Peletier R.F. 2001, AA, 367, 428

 [4] Aguerri J.A.L., M\'endez-Abreu J., Corsini E.M. 2009, AA, 495, 491

 [5] Athanassoula E., 2003, MNRAS, 341, 1179

 [6] Barnes J.E., Hernquist L. 1992, ARA$\&$A, 30, 705

 [7] Athanassoula E., Beaton R.L. 2006, MNRAS, 370, 1499

 [8] Bosma A. 1983, in Athanassoula E., ed., Proc. IAU Symp. 100, Internal Kinematics and Dynamics of Galaxies, Reidel,
   Dordrecht, p. 253

 [10] Bedregal A., Arag\'on-Salamanca A., Merrifield M. 2006, MNRAS, 373, 1125 

 [11] Bekki K., Couch W.J. 2011, MNRAS, 415, 1783

 [12] Bender R. 1988, Astr. astrophys Lett, 202, L5

 [13] Bender R., Surma P., Dobereiner S., M\'ollenhoff C., Madeijsky R. 1989, AA, 217, 35

 [14] Bournaud F.,  Combes F. 2002, AA, 392, 83

 [15] Buta R., Crocker D.A. 1991, AJ, 102, 1715

 [16] Buta R. 1995, ApJS, 96, 39

 [17] Buta R., Vasylyev S., Salo H., Laurikainen E. 2005, AJ, 130, 506

 [18] Buta R., Laurikainen E., Salo H., Block D.L., Knapen J.H. 2006, AJ, 132, 1859

 [19] Buta R. J., Corwin H. G., Odewahn S. C. 2007, The de Vaucouleurs Atlas
   of Galaxies, Cambridge, Cambridge University Press

 [21] Buta R., 2011, to appear in 'Planets, Stars, and Stellar Systems', Vol. 6, Series Editor T. D. 
   Oswalt, Volume editor W. C. Keel, Springer 
    1102.0550

 [22] Cappellari M. et al. 2011, MNRAS, 416, 1680

 [23] Combes F., Debbasch F., Friedli D., Pfenniger D. 1990, AA, 233, 82

 [24] Combes F., Sanders R.H. 1981, AA, 96, 164

 [25] Djorgovski S. 1992, in ``Morphological and Physical Classification of
   Galaxies'', eds. Longo G., Capaccioli M., Busarello G.. Dordrecht: Kluwer Academic Publishers, 427

 [26] Dressler A. 1980, ApJ, 236, 351

 [27] Dressler A., Sandage A. 1983, ApJ, 265, 664

 [28] Elmegreen B., Elmegreen D. 1985, ApJ, 288, 438

 [29] Emsellem E. et al. 2011, MNRAS, 414, 888

 [30] Erwin, P., Beltr\'an J. C., Graham A. W., Beckman J.E. 2003, ApJ, 597, 929

 [31] Eskridge P., Frogel J.A., Pogge, R.W., Quillen A.C., Berlind A.A., Davies R.L., DePoy D.L., 
     Gilbert K.M., Houdashelt M.L., Kuchinski L.E.+5 co-authors 2002, ApJS, 143, 73

 [32] Freeman K.C. 1970, ApJ, 160, 811

 [33] Gadotti D. 2009, MNRAS, 393, 1531

 [34] Graham A., Guzm\'an R. 2003, AJ, 125, 2936

 [35] Graham A., Worley C. 2008, MNRAS, 388, 1708

 [36] Gunn J.E., Gott J.R.I. 1972, ApJ, 176,1

 [37] Hubble 1936, {\it Realm of Nebulae}. New Haven; Yale Univ. Press

 [38] Icke V. 1985, AA, 144, 115

 [39] Khochfar S., Burkert A. 2005, MNRAS, 359, 1379

 [40] Khochfar S., Silk J. 2006, ApJL, 648, L21

 [41] Khochfar S. et al.  2011, MNRAS, 417, 485

 [42] King I., 1992, in ``Morphological and Physical Classification of
     Galaxies'', eds. Longo G., Capaccioli M., Busarello G.. Dordrecht: Kluwer Academic Publishers

 [43] Kormendy J. 1979, ApJ, 227, 714

 [44] Kormendy J., Bender R. 1996, ApJ, 464, L119

 [45] Kormendy J., Kennicutt R. 2004, ARA$\&$A, 42, 603

 [46] Larson R.B., Tinsley B.M., Caldwell C.N. 1980, ApJ, 237, 692

 [47] Laurikainen E., Salo H. 2002, MNRAS, 337, 1233

 [48] Laurikainen E., Salo H., Buta R. 2004, ApJ, 607, 103 

 [49] Laurikainen E., Salo H., Buta R. 2005, MNRAS, 362, 1319

 [50] Laurikainen E., Salo H., Buta R., Knapen J., Speltincx T., Block D.L. 2006, AJ, 132, 2634

 [51] Laurikainen E., Salo H., Buta R., Knapen J. H. 2007, MNRAS, 381, 401

 [52] Laurikainen E., Salo H., Buta R., Knapen J. H. 2009, ApJL, 692, L34

 [53] Laurikainen E., Salo H., Buta R., Knapen J. H., Comerón S. 2010, MNRAS, 405, 1089

 [54] Laurikainen E., Salo H., Buta R., Knapen J. H. 2011, MNRAS, 418, 1452

 [55] Martinez-Valpuesta I., Knapen J.H., Buta R. 2007, AJ, 134, 1863

 [56] Martig M., Bournaud F. 2010, ApJ, 714, 275

 [57] Moore B., Katz N., Lake G., Dressler A., Oemler A. 1996, Nat. 379, 613 

 [58] Naab T., Trujillo I. 2006, MNRAS, 369, 625

 [59] Naab T., Khochfar S., Burkert A. 2006, ApJ, 636, 81

 [60] Nair P.B., van den Berg S., Abraham R.G. 2010, ApJ, 715, 606

 [61] Guo Q., White S., Boylan-Kolchin M., De Lucia G., Kauffmann G., Lemson G., Li C., 
    Springel V., Weinmann S. 2011, MNRAS, 413, 101

 [62] Salo H., Rautiainen P., Buta R., Purcell G.B., Cobb M.L., Crocker D.A., Laurikainen E. 1999, AJ, 117, 792

 [63] Salo H., Laurikainen E., Buta R. 2004, ASSL, 319, 673

 [64] Salo H., Laurikainen E., Buta R., Knapen J. 2010, ApJ, 715, 56

 [65] Scannapieco C., White S.D.M., Springel V., Tissera P.B. 2011, MNRAS, 417, 154

 [66] Scannapieco C., Tissera P.B. 2003, MNRAS, 338, 880

 [67] Simien F., de Vaulouleurs G. 1986, ApJ, 302, 564

 [68] Toomre A., Toomre J. 1972, ApJ, 178, 623

 [69] Thomas D., Maraston C., Bender R., Mendes de Oliveira C. 2005, ApJ, 621, 673
 [71] Spitzer L. Jr., Baade W. 1951, ApJ, 113, 413

 [72] van den Bergh S. 1976, ApJ, 206, 883

 [73] de Vaucouleurs G., de Vaucouleurs A., Corwin H. G., Buta R., Paturel G., 
    Fouqu\'e, P. 1991, Third Reference Catalogue of Bright Galaxies, New York, Springer (RC3)

 [74] Zhang X. 1996, ApJ, 457, 125

 [75] Laine S., Shlosman I., Knapen J.H., Peletier R.F. 2002, ApJ, 567, 97

 [76] Erwin P., Sparke L. 2003, ApJS, 146, 229

 [77] Athanassoula E., Sellwood J.A. 1983, in Athanassoula E., ed., Proc. IAU Symp. 100, Internal Kinematics and Dynamics of Galaxies, Reidel,
   Dordrecht, p. 203
 
 [78] Balcells M., Graham A.W., Dom\'inguez-palmero L., Peletier R.F. 2003, ApJ, 582, 79

 [79] Athanassoula E., Misioritis A. 2002, MNRAS, 330, 35 

 [80] Block D.L., Bournaud F., Combes F., Puerari I., Buta R. 2002, AA, 394, L35

 [81] Weinzirl T., Jogee S., Khochfar S., Burkert A., Kormendy J. 2009, ApJ, 696, 411

 [82] Aguerri J.A.L., Elias-Rosa N., Corsini E.M., Munoz-Tun\'on C. 2005, AA, 434, 109

 [83] Gadotti D. 2008, MNRAS, 384, 420

 [84] Balcells M., Graham A., Peletier R.F. 2007, ApJ, 665, 1104

 [85] Laurikainen E., Salo H. 2001, MNRAS, 324, 685

 [86] Buta R., Laurikainen E., Salo H., Knapen J.H. 2010, ApJ, 721, 259

 [87] Buta R., Laurikainen E., Salo H. 2004, AJ, 127, 279

 [88] Kormendy J., Bender R. 2011 (astro-ph:1110.4384) 

 [89] Burstein D., Ho L., Huchra J.P., Macri L.M. 2005, ApJ, 621, 246

 [90] Geach J.E., Smail I., Moran S.M., Treu T., Ellis R. 2009, ApJ, 691, 783

 [91] Laurikainen E., Salo H., Rautiainen P. 2002, MNRAS, 331, 880

\end{document}